# Graphene Bilayers with a Twist

Eva Y. Andrei[1*]   and Allan H. MacDonald[2]

[1]Department of Physics and Astronomy, Rutgers the State University, Piscataway, NJ
[2]Department of Physics, The University of Texas at Austin, Austin, TX
*Corresponding author

**Abstract**

**Near a magic twist angle, bilayer graphene transforms from a weakly correlated Fermi liquid to a strongly correlated two-dimensional electron system with properties that are extraordinarily sensitive to carrier density and to controllable environmental factors such as the proximity of nearby gates and twist-angle variation. Among other phenomena magic-angle twisted bilayer graphene hosts superconductivity, interaction induced insulating states, magnetism, electronic nematicity, linear-in-T low-temperature resistivity, and quantized anomalous Hall states. We highlight some key research results in this field, point to important questions that remain open, and comment on the place of magic angle twisted bilayer graphene in the strongly correlated quantum matter world.**

**Main**

The two-dimensional materials field can be traced back to the successful isolation of single-layer graphene sheets exfoliated from bulk graphite[1]. Galvanized by this breakthrough, researchers have by now isolated and studied dozens of 2D crystals[2] and predicted the existence of many more[3]. Because all the atoms in any 2D crystal are exposed to our three-dimensional world, this advance has made it possible to tune the electronic properties of a material without changing its chemical composition, for example by introducing a large strain[4-10], plucking out atoms[11-13], stacking 2D crystals[14,15], or simply by using electrical gates to add or remove electrons. One of the simplest techniques, changing the relative orientation between 2D crystals[14-19] has been especially impactful following the discovery of interaction induced insulating states and superconductivity in twisted bilayer graphene with a special *magic* twist angle (TBG)[20-26]. In this review we survey some of the progress that has been made toward understanding these phenomena. Our coverage is necessarily incomplete due to both limitations in space and the rapidly evolving research landscape. We limit our attention to TBG, although twist-angle control is now also proving its value in other 2D crystal systems[27-33].

**Energy bands of an isolated graphene sheet**
We first consider monolayer graphene. Its electronic properties derive strictly from geometry: a 2D honeycomb lattice with identical circularly symmetric orbitals on interpenetrating triangular Bravais sub-lattices, labeled A and B (Figure 1a), and identical nearest neighbor hopping parameters *t*. These conditions lead to a unique band structure that at low energy, and in the absence of interactions, is described by a one-parameter 2D Dirac-Weyl Hamiltonian:

$$H = v_F \begin{pmatrix} \sigma \cdot p_K & 0 \\ 0 & -\sigma^* \cdot p_{K'} \end{pmatrix}$$

where $v_F = \frac{3}{2\hbar} at$ is the Fermi (or Dirac) velocity, $\hbar$ the reduced Planck constant, $a = 0.142$ nm is the Carbon-Carbon distance, t ~ 2.7 eV is the hopping parameter, $\sigma = (\sigma_x, \sigma_y)$ are the Pauli matrixes operating on the sub-lattice degree of freedom, $p_K = \hbar(k - K)$ and $p_{K'} = \hbar(k - K')$ is the momentum measured relative to the $K$ and $K'$ corners ($K = \frac{4\pi}{3a_0}$) of the hexagonal Brillouin zone (BZ) of graphene (Figure 1b) respectively, and $a_0 = \sqrt{3}a$ is graphene's lattice constant. The low energy band structure consists of a pair of inequivalent Dirac cones (linear crossings between the conduction and valence bands), shown in Figure 1c. The apexes of the Dirac cones, referred to as Dirac points, are located at the three equivalent K corners (full circles in Figure 1b) or K' corners (open circles in Figure 1b) of the BZ. Since $\sigma \cdot p_K$ is the chirality operator, and $-\sigma^* \cdot p_{K'} = \mathcal{T}(\sigma \cdot p_K)$ is its time reversal partner ($\mathcal{T}$ is the time reversal operator), states in the $K$ and $K'$ Dirac cones have opposite chirality. $K$ and $K'$ define a valley degree of freedom, which together with spin produces a 4-fold band degeneracy. The Dirac points separate the conduction and valence bands so that at charge-neutrality the Fermi surface (pink plane in Figure 1c) consists of just these two points. Owing to the conical energy-momentum relationships, the electron and hole-like quasiparticles move at a constant speed, $v_F$ ~ $10^6$ m/s, which is determined by the slope of the cone[34-36]. As a result, the charge-carriers behave as if they were massless ultra-relativistic particles[34], albeit with a non-universal velocity that is approximately 300 times smaller than the speed of light. The band crossings defining the Dirac points are protected by three discrete symmetries[37]: $C_{2z}$ (in-plane inversion), $\mathcal{T}$, and $C_3$ (three-fold rotation), which prevent a gap from opening in graphene's spectrum unless one of the symmetries is broken. The absence of a gap stands as an obstacle to applications of graphene that require switching, such as transistors. The conical band structure produces a density of states (DOS) which, at low energies, is linear in energy, $E$, and vanishes at the Dirac point, $\rho(E) \sim \frac{2|E|}{\pi(\hbar v_F)^2}$, making it easy to change the Fermi energy by electrostatic doping. Because the band structure of graphene is a direct consequence of its geometry, graphene-like band-structures have also been realized in other systems satisfying these conditions, including artificial structures and cold atoms[38-40].

**Twisted bilayer graphene: moiré superlattices and electronic properties**
Moiré patterns created by superimposing 2D meshes have a long history in textiles, art, and mathematics. In surface science moiré patterns made their debut shortly after the invention of the scanning tunneling microscope (STM) which enabled the observation of anomalously large super-periodic patterns on the surface of graphite[41]. More than two decades elapsed before scientists observed that, beyond having aesthetic appeal, moiré patterns can drastically alter electronic properties[14]. The STM moiré pattern obtained by superposing two graphene crystals with a twist-angle $\theta$ between the two layers consists of an array of alternating bright and dark spots with period, $L \approx a_0/(2\sin(\theta/2))$ (Figures 1d and 2a). The bright spots correspond to regions referred to as AA, where every atom in the top layer has a partner directly underneath it in the bottom layer. In the darker regions the local stacking is approximately A on B or B on A, referred to as AB or BA respectively, and also known as Bernal stacking. In the AB case, each top-layer A atom sits directly above an atom in the bottom layer while top-layer B atoms have no partner in the bottom layer, and vice versa for BA stacking.

The large moiré period leads to a moiré Brillouin-zone that consists of two inequivalent hexagons of side, $k_\theta \sim 2K\sin(\theta/2)$. These mini-BZs are spanned by the shift between the K (K') corners of the of the BZs of the top (1) and bottom (2) layers, $\mathbf{K^1} - \mathbf{K^2}$ and $\mathbf{K'^1} - \mathbf{K'^2}$ respectively (Figure 1e). States in the K and K' mini-BZs have opposite chirality inherited from their parent Dirac cones. The low energy Hamiltonian of TBG for the K mini-BZ is given by:

$$H_K = \begin{pmatrix} v_F \boldsymbol{\sigma}.\mathbf{p}_1 & T(\mathbf{r}) \\ T^+(\mathbf{r}) & v_F \boldsymbol{\sigma}.\mathbf{p}_2 \end{pmatrix}$$

where $\mathbf{p}_i = \hbar(\mathbf{k} - \mathbf{K}_i)$, $i = 1,2$ corresponds to the top and bottom layer respectively, and the Hamiltonian for the K' mini-BZ is its time reversal partner. $T(\mathbf{r})$ is the sub-lattice dependent moiré potential that couples the two layers[42,43]. The higher energy of AA stacking compared to that of Bernal stacking, about 20meV/atom, modulates the interlayer distance[44] by ~0.015 nm and consequently also the interlayer tunneling strength. In the absence of interlayer tunneling, the band structure consists of four Dirac cones from the two layers and two valleys that intersect at two energies separated by $\Delta E = \pm \hbar v_F k_\theta$ (Figure 1f). Turning on tunneling between the layers produces avoided crossings at these intersections leading to saddle points in the band structure. Saddle points are points in momentum space at which an energy band reaches energy minima and maxima along orthogonal directions in momentum space. In 2D systems saddle points create dramatically enhanced DOS peaks that are easily identified in scanning tunneling spectroscopy (STS) studies[14], and referred to as van Hove singularities (VHS). At large $\theta$, the Dirac cones are widely separated, and the low energy states in one layer are only weakly influenced by tunnel coupling to the adjacent layer. The energy separation between the conduction and valence band VHSs at large twist angles, $\Delta E \sim \hbar v_F k_\theta - 2w$, is equal to twice the isolated layer energies at the Dirac cone intersection points, less an avoided-crossing inter-layer tunneling energy shift. An estimate of the interlayer tunneling strength can be obtained from Bernal stacked bilayer graphene, $w \sim 0.1$eV. For $\theta > 10^0$, where $\Delta E > 1$eV, the low energy electronic properties of the TBG are nearly indistinguishable from those of isolated graphene[14,16] layers. Still, the presence of the second graphene layer has an influence, since it efficiently screens random potential fluctuations introduced by the substrate[45], allowing access to low energy electronic properties that would otherwise be obscured by the substrate-induced potential fluctuations[46]. The same screening effects also acts on electron-electron interactions, of course, and may have the tendency to suppress strong correlation physics. When the twist-angle is reduced, the VHSs come closer together, tunneling between layers couples the low energy Dirac cone states, the hybridization between layers becomes strong, and the bilayer can no longer be viewed as consisting of two weakly-coupled layers. All electronic wavefunctions are then coherent superpositions of components on the four sublattices of the bilayer with intricate correlated position dependences.

The unusual electronic properties of TBG were revealed early on by STS on graphene layers synthesized by chemical vapor deposition (CVD)[14]. STS, which is the spectroscopic counterpart of STM, monitors the differential conductance, dI/dV, with respect to energy (or bias voltage) and is approximately proportional to the local electronic DOS at the tip position. The CVD samples, which were supported on a Au coated TEM (transmission electron microscope) grid, contained large regions with uniform moiré patterns (Figure 2a) indicating extended domains with uniform twist-angle. Surprisingly at the time, the STS spectra on TBG samples were completely different from those on either single layer or bilayer (Bernal-stacked) graphene, featuring a pair of pronounced VHS peaks that flanked the minimum DOS marking the Dirac point (Figure 2b bottom trace). The energy separation of the VHS peaks decreased monotonically with twist-angle, $\theta$. Near $\theta_0 \approx \frac{3}{2\pi}\frac{w}{t} \approx 1.1^o$, now recognized as the *magic* angle[47], the two peaks merged into a narrow, ~ 32mV wide peak now recognized as reflecting the formation of flat-bands (Figure 2b top trace). The observation of a 12mV pseudo-gap feature at the Fermi energy and the concomitant appearance of a spatial modulation in the DOS maps, suggested the emergence of an interaction driven insulating or reduced conductivity state, which was interpreted at the time as evidence for a charge density wave. Subsequent Landau level spectroscopy measurements on these samples revealed that in addition to introducing VHSs, the moiré patterns also lowered the Fermi velocity[16] (Figure 2c, 2d), as predicted theoretically[47].

The observation of VHSs in TBG stimulated a large body of both theoretical and experimental work[47-56]. It was understood that the VHS peaks, one in the valence band and one in the conduction band, arose due to saddle points in the energy bands. The evolution[19] of electronic properties with twist-angle was fully explained by a microscopic tight-binding model, and also by a simple and physically transparent model[47] in which the moiré pattern is encoded in the variation of sublattice-dependent interlayer tunneling with local stacking arrangement. A sequence of magic twist angles was found, the largest being $\theta_0$, where the highest energy valence miniband and the lowest energy conduction miniband simultaneously approach a common energy throughout the moiré superlattice BZ, yielding vanishing Fermi velocities (Figures 2c, 2d) and two extremely flat-bands (Figure 2e). In this work, the same value of the interlayer tunneling strength $w$, was assumed for AA and AB sites. It was later found that in the limit of vanishing AA tunneling rate, the magic twist angle sequence follows a simple analytic expression[42,57], $\theta_n \approx \theta_0 \frac{\alpha_0}{\alpha_n}$, where $\alpha_0 = \frac{w}{\hbar v_F K \theta_0} \approx \frac{1}{\sqrt{3}}$, $\alpha_n \approx \alpha_0 + 3/2n$.

The reduction in band energy scales near magic twist-angles suggests many-body ground states that are controlled dominantly by electronic correlations[18,19,47]. These electronic structure models were later verified and refined[58-60] by fully self-consistent *ab initio* density-functional theory (DFT) electronic structure calculations that importantly, also account for the strain-pattern of the moiré superlattice[60]. The *ab initio* calculations are cumbersome because of the large number of carbon atoms (~ 13,000 C) per moiré period in magic-angle TBG (MATBG), and this motivates the use of alternative approaches that provide more economical descriptions. At the magic-angle almost all band dispersion is lost, but what remains is sensitive to the approximation scheme employed[61,62]. Independent of these details, the lowest energy conduction and valence bands are always very flat over most of the superlattice Brillouin zone. At larger twist angles the bands are more dispersive and less sensitive to details, recovering the V-shaped pseudo gap-feature of isolated graphene.

The low energy band structure of MATBG features 4 weakly dispersive (flat), spin-degenerate, topological bands[43,63], two each in the hole and electron sectors, that are isolated from higher energy dispersive bands by energy gaps throughout the two mini-Brillouin-zones[60]. Each mini-BZ hosts two Dirac cones, one from each layer, which are located at its corners and, as in an isolated graphene sheet, protected by a $C_{2z}\mathcal{T}$ symmetry. Breaking this symmetry, for example by aligning one of the graphene layers with the hBN to break $C_{2z}$, or by applying a magnetic field to break $\mathcal{T}$, opens spectral gaps[64-68].

Because the band-filling can be adjusted with electrical gates, the many-electron ground state of MATBG can be studied experimentally over the full range of band-fillings without adding chemical dopants[20-22,64-70]. The band-filling factor, $\nu$, defined as the number of carriers per moiré cell, ranges from -4 when all the bands are empty, to 0 when half of the bands are filled and the sample is electrically neutral, to +4 when all the bands are filled. $\nu = \pm 1, \pm 2, \pm 3$, then corresponds to 1,2,3, electrons (+) or holes (-) per moiré cell, and 0 indicates the charge-neutrality point. The flat-bands are sensitive to strain and can be broadened significantly by twist-angle inhomogeneity[46,57] and other sources of disorder. They also broaden considerably and depend strongly on the Fermi level when electrostatic and exchange interaction effects are properly taken into account[71,72]. The quasiparticle lifetimes in these flat-band states are expected to be short due to strong interaction effects, discussed further below.

**Superconductivity in tear and stack MATBG revealed by transport measurements.**

Although naturally grown CVD TBG provided hints of correlated behavior, the inability to control twist-angle made it difficult to gain deeper insight. Subsequent developments were enabled by encapsulation in hBN (hexagonal boron nitride) as protection against external disturbances[73,74] and by a breakthrough in sample fabrication, the tear-and-stack method[15] which enabled precise twist-angle control. Taking advantage of the stronger adhesion of graphene to hBN than to $SiO_2$, a graphene flake deposited on $SiO_2$ is cut in two by pressing an hBN flake on one half of it and lifting it up. The segment that was not contacted by the hBN is left on the substrate which is then rotated by the desired twist angle before being picked up by the graphene-hBN stack. In order to carry out transport or scanning tunneling spectroscopy measurements, additional fabrication steps are necessary, including depositing the stack on another hBN flake to complete the encapsulation and depositing metallic gates and contacts. These technical breakthroughs made it possible to achieve high quality MATBG with precisely controlled twist-angles.

Low-temperature (~1K) electrical transport measurements on high quality tear-and-stack MATBG protected from the environment by encapsulation in hBN (Figure 3a) revealed interaction-induced insulating states at integer moiré band fillings (Figure 3d) together with nearby superconducting domes[20,22] (Figure 3b). The similarity between the phase diagram and phenomenology of this relatively simple system and that of the high-temperature superconductors (HTS), prompted a flurry of activity in search of clues that might help answer long-standing HTS questions[62,75-85]. A prime advantage of MATBG over the HTS compounds, in which changing the doping levels typically requires chemical synthesis of a different sample, is that in MATBG the doping-dependence can be accessed by using electrostatic gating to tune the carrier density of an individual device.

**The ubiquity of twist-angle disorder in tear-and-stack samples**

It is known from Raman spectroscopy[86], AFM and STM[23-26] (Figure 4a, 4b), that tear-and-stack samples can have sizable twist-angle inhomogeneity over the active area of the device that depends on details of the fabrication procedure. Since the observed properties vary sharply with moiré-filling, which for a fixed electron density is inversely proportional to the square of the twist angle, even small twist-angle variations lead to spatial inhomogeneity in physical properties, especially at higher band-filling. For example for $\nu = 3$, even a 5% variation in twist-angle will change the band-filling by ~ 0.3, and therefore acts as a strong source of disorder. Since, as shown above, the value of the magic twist-angle, $\theta_0$, depends on the ratio of the interlayer to intralayer tunneling strength, $w/t$, applying pressure[87,88] which increases $w$, increases the magic twist-angle value and provides some degree of tuning capability that can reduce the severity of absolute twist-angle control requirements.

Twist angle inhomogeneity is thought to be introduced during sample fabrication by small variations in local strain, blisters, or substrate imperfections. Large scale maps obtained with a scanning SQUID (superconducting quantum interference device)[89], revealed that even TBG samples that exhibit superconducting domes contain local twist-angle inhomogeneity (Figure 4c) which can be as large as 10% of the magic twist-angle. Inhomogeneity in tear-and-stack samples can be mitigated by squeezing out the blisters and folds during sample fabrication[90], but this comes at the cost of a less reliable overall twist-angle outcome. One advantage of naturally grown TBG is that the twist-angle is homogeneous over much larger areas (Figure 2a) without the need for any special treatment.

**Flat band MATBG observed with local probes**

Inhomogeneity in twist-angle is particularly deleterious for measurements, such as electrical-transport, that probe the entire sample. For example, if magic-angle regions do not form a contiguous path connecting the source and drain electrodes, superconductivity may not be observed. Similarly, twist-angle inhomogeneity will suppress the magnitude of insulating gaps inferred from Arrhenius plots of the temperature dependent resistivity. In the presence of sample inhomogeneity, access to the electronic properties can still be gained by using local probes such as STM/STS.

STM enables direct visualization of the atomic-structure, the local twist-angle and its homogeneity, domain walls, and lattice disorder. STS provides access to the electronic structure, energy gaps and their position relative to the Fermi level, as for example in superconductors or in 2D electronic-systems in magnetic-fields with discrete Landau levels[35,91,92]. It also enables monitoring the response to perturbations such as strain[6], point defects[93-95], doping[96], and magnetic field[91]. Importantly, STS is able to interrogate electronic properties in a large energy window both above and below the Fermi energy. This is in contrast to electrical-transport which mainly probes the electronic response of the carriers at the Fermi-energy, and also differs from angular-resolved photo-emission spectroscopy (ARPES) which probes the band-structure below the Fermi-energy. The STS response is never completely local however, given the non-locality of quantum mechanics, and measurements are still influenced to some degree by inhomogeneity away from the tip position. Among the challenges of STM/STS is the lack of optical access in most standard probes, which makes it difficult to locate the micron size MATBG samples. This obstacle has been overcome by employing a technique that uses the STM tip as a capacitive antenna, acting as a GPS-like locator[97].

The flat-band DOS peak in MATBG revealed by STM/STS measurements on both naturally grown and tear-and-stack MATBG is concentrated in the AA stacked regions of the moiré-cell[14,23-26], consistent with theoretical predictions [51,98]. The shape, structure, and width of the DOS peak, are particularly sensitive to whether or not the flat-band is partially filled. When the Fermi-level is outside the peak, corresponding to either an empty or full band, the experimental linewidth is at its minimum[23-26,99] (Figure 4d). The structure of the flat-band DOS peaks in this case vary significantly from sample to sample, ranging from a single Gaussian-shaped peak[24,99] to two peaks[23,25,26]. These discrepancies may reflect broadening due to intrinsic effects such as differences in the strength of Coulomb interactions[72] or lattice relaxation[60], as well as extrinsic ones such as disorder, twist-angle heterogeneity[61], strain[100,101], and sublattice symmetry broken by the hBN substrate. The flat-band peak is flanked by two minima displaced by ~50 mV (~75mV) from its center on the hole (electron) side, marking the energy gaps that isolate it from the closest remote bands, consistent with band structure calculations for relaxed lattices[60]. Doping the sample to partially fill the flat band, completely reconstructs the peak by introducing a strong pseudogap feature at the Fermi level[23-26]. Furthermore, measuring the filling dependence of the differential conductance at the Fermi level, which is proportional to the local DOS, reveals pronounced minima at all integer fillings of the flat band (Figure 4e). This indicates the emergence of incipient correlated insulator states, which mirror the conductance minima at integer fillings observed in transport measurements[20,22,64-66] (Figure 3d). Similarly, local inverse compressibility measurements, which are proportional to the inverse DOS, revealed a sequence of peaks at integer fillings[69], consistent with jumps in chemical potential across a gap (Figure 4f). As far as we know, in all measurements reported to date these conductivity, or DOS, minima tend to be much more pronounced in the electron sector than in the hole sector, possibly reflecting a larger gap isolating the flat bands on the conduction band side.

**Similarities between MATBG and the high-temperature superconductors**
Interaction-induced insulators and superconducting domes. The basic phenomenology of MATBG can be summarized by a phase-diagram in which superconducting domes flank Mott-like insulating phases (Figure 3b), resembling the phenomenology of cuprate[102-105] (Figure 3c) and pnictide[106,107] superconductors.

Strong coupling. In MATBG the ratio of the superconducting transition temperatures to band widths or Fermi temperatures, $T_c/T_F \sim 0.1$, exceeds the range at which the weak coupling theory of superconductivity can be employed, and is similar to $T_c/T_F$ in other materials (Figure 3e) that exhibit superconductivity close to metal-insulator phase transitions[21].

Pseudogap state. Cuprates and pnictides feature a pseudogap state above the superconducting dome[103]. In this state STS measurements have shown that the DOS is suppressed without being fully gapped[108] consistent with a spectral weight depletion observed in ARPES[109,110]. In MATBG, STM/STS measurements at temperatures above the superconducting transition in the AA stacked regions of the moiré-cell showed that, similar to observations in HTS, a pseudogap opens at the Fermi energy[23-26] in the partially filled flat-band (Figure 4d).

Strange metal phase. A common thread among strongly correlated superconductors including cuprates, pnictides, organic-superconductors, and heavy fermions[111-113] is that their high temperature metallic parent states are strange metals with anomalous features that cannot be described in terms of the coherent quasiparticle excitations that underpin conventional Fermi-liquids. One manifestation of this state is linear in temperature resistivity that persists down to temperatures well below the Debye scale. In this regime it is claimed that many dissimilar materials exhibit a universal, Planckian, behavior[114,115], that is characterized by a resistively determined scattering rate, $\Gamma = c \frac{k_B T}{\hbar}$ where $c \sim 1$. This universal behavior, which suggests a quantum limit on the scattering rate, has been associated with quantum criticality in strongly correlated systems[116]. Resistivity measurements[101,117] in MATBG doped within the flat-band, revealed a linear temperature dependence over a wide range of band filling, suggesting the existence of a strange-metal phase above the superconducting domes. The scattering rates extracted from these measurements at magic angle twists were claimed to be consistent with the Planckian limit[101], providing another link between the phenomenology of MATBG and that of strongly correlated electronic systems near a quantum critical transition. An alternative interpretation of the linear in T resistivity results in terms of phonon scattering has not however been ruled out[84].

Broken symmetry states in the pseudogap and superconducting phase. The pseudogap phase in HTS compounds often hosts nematic, charge-density wave, and spin-density wave broken symmetry states that can compete or coexist with the superconducting phase[104-107]. Similar phenomena were recently uncovered in MATBG probed with STS (Figure 5a), and transport measurements (Figure 5c). DOS maps obtained in STS have shown that when the flat-band is partially filled, the rotational symmetry of the moiré cells is broken[23,24]. The charge redistribution in the nematic state, which was obtained with an STS based local charge spectroscopy technique[24], revealed a global nematically ordered state with alternating charge stripes. The nematic charge order, which is present only in the partially filled flat bands of MATBG, survives to temperatures as high as T~35K and to magnetic fields B ~ 8T[24]. These findings reinforce the idea that nematic order which redistributes the charge within the moiré-cell is an intrinsic property of the correlated states in the partially filled flat-band[118], and that it may be an important precursor to the superconducting states that emerge at lower temperatures. Magneto-transport measurements in MATBG found evidence of spontaneously broken lattice rotational symmetry in both the normal and superconducting phases[119]. In these measurements the nematic order was inferred from the observation of resistivity anisotropy above the superconducting transition. In the superconducting phase, the appearance of an anisotropic in-plane critical field coupled with an anisotropic superconducting critical current suggested that the superconducting state is also nematic.

Reduction of superconducting transition and competing orders. In certain underdoped cuprates, a dip in the critical temperature that defines the superconducting dome at around 1/8 hole doping has been attributed to a competition between superconductivity and spin or charge ordered nematic phases[104,105]. Transport measurements in MATBG revealed a similar reduction of the critical temperature at the crossing between the anisotropic normal state and the superconducting dome on the hole side of the insulator at ν = -2, (Figure 5c) again resembling the phenomenology of cuprate superconductors[119].

**More surprises**
The flat-band correlation physics[47] revealed by experiment, especially the important role played by superconductivity, has been full of surprises. The resemblance between the phenomenology of HTS and MATBG is quite striking, but whether these similarities reflect a deep connection remains to be seen. With many new results pouring in, it now appears that in some respects the physics of the MATBG is even richer than that of cuprates.

We first discuss the superconducting domes. Whereas in the high temperature superconductors (HTS) there are typically two superconducting domes flanking an antiferromagnetic insulating state, in the MATBG case, there are 7 intermediate integer fillings between full and empty at which insulating states can appear, and correspondingly many non-integer filling intervals in which superconducting domes can potentially appear. Transport measurements in samples treated to reduce inhomogeneity[22] showed a gap at the charge-neutrality point and provided evidence that insulating states likely occur at all integer fillings in ideal samples, and that superconducting domes can occur very close to charge-neutrality in addition to domes close to other integer fillings (Figure 5b). The phase diagram of MATBG is sensitive to the proximity of the gates used to vary carrier density[132], suggesting that weaker Coulomb interactions favor superconducting states over insulating states.

Orbital magnetism is another property unique to MATBG. In samples in which one of the graphene layers was aligned with the hBN encapsulant, a large ~ 6 meV gap was observed at charge-neutrality, likely because the hBN substrate[120] breaks $C_{2z}$ sublattice symmetry and opens gaps at the Dirac points in the mini-Brillouin zones. In the presence of an out of plane magnetic field, a hysteretic Hall signal[64,65] was observed at moiré filling ν = 3. When the field was swept through zero, the Hall resistance was quantized to within 0.1% of the von Klitzing constant $h/e^2$, reflecting a topological state with Chern number C = 1 that is an orbital ferromagnet.[121-123]. The state observed at ν = 3 is thought to be a spin and valley polarized ferromagnetic Chern insulator which is, like many gapped states in graphene Landau levels, stabilized by exchange interactions[42,43,123-127]. The topological nature of the bands has been robustly revealed even without alignment to the hBN substrate[66-68], by breaking the time reversal component of the $C_{2z}\mathcal{T}$ symmetry with an external magnetic field. Hall density measurements revealed van Hove singularities leading to Lifshitz transitions[66] (Figure 5d) and to Chern insulators with well-developed quantized Hall resistance plateaus corresponding to Chern numbers C=±1, ±2, ±3 at fillings ν = ∓3, ∓2, ∓1 respectively (Figure 5e).

Because graphene sheets have negligible spin-orbit coupling[128], the orbital and spin contributions to the magnetization of graphene-based Chern insulators can be estimated separately. The spin-magnetization at zero temperature is especially simple because it results from a fully spin-polarized moiré superlattice miniband that contributes one Bohr magneton per superlattice unit cell. The scale of the orbital magnetization can be estimated using a special property of Chern insulators, namely that its orbital magnetization varies linearly with chemical potential, even when the chemical potential lies in the gap of the insulator. It follows that the difference in orbital magnetization per unit cell between a weakly n-doped insulator and a weakly p-doped insulator is

$$\frac{\Delta M A_M}{A \mu_B} = C \frac{m E_g A_M}{\pi \hbar^2}$$

where $C$ is the integer-valued topological Chern index of the insulator, $A_M$ is the moiré unit cell area, A the sample area, $E_g$ is the energy gap, and $m$ is the electron mass[122]. Because $A_M$ exceeds the microscopic unit cell area by a factor of more than $10^4$ at the magic-angle, the right-hand side of this expression is larger than one, even for gaps on the meV scale. The magnetization is dominantly orbital, even more so when thermal fluctuations and very weak anisotropy in the spin magnetism are taken into account. The Chern insulators of MATBG, as far as we are aware, provide the first example of magnetism in which time-reversal symmetry is broken independently in dominant orbital degrees of freedom. The very strong dependence of the magnetization on doping[129] is a unique property of orbital ferromagnets.

**Theoretical progress**
A rather large body of theoretical work has been undertaken to understand the properties of MATBG. We now have a relatively good understanding of the mathematical origin of the surprising flat band property. In graphene bilayers flat bands result from interference between inter and intra-layer tunneling, and not from simple electron-localization behavior. Qualitatively, the flat band behavior can be understood[43,63] in terms of effective magnetic fields that act on electrons near AA positions in the moiré pattern and are produced by interlayer electron hopping. When tunneling between layers on the same sublattice is neglected, it can be rigorously[42] shown that a pair of perfectly flat isolated bands appear at a discrete set of twist angles, and that their wavefunctions are reminiscent of Landau level wavefunctions on a torus.[43,63] These developments suggest[130] some similarity between the interaction physics in MATBG and the interaction physics that is responsible for the fractional quantum Hall effect. Indeed, the phenomenology of gapped incompressible states at integer filling factors in MATBG, which require interactions because of the spin and flavor band degeneracies, is reminiscent of Landau level physics in graphene and is the second aspect of MATBG phenomenology that appears to be relatively well understood. Just as in the case of graphene Landau levels, interaction-induced insulating states are formed in the flat bands at non-zero integer moiré band filling factors by breaking spin/valley flavor symmetries to reduce interaction energy. Because the flavor projected bands are normally topological, this picture also provides a natural explanation for the discovery of the quantum anomalous Hall effect in MATBG discussed above. The gate voltage dependence of the weak-field magnetic oscillations that are apparent in transport measurements[46] suggest that these broken symmetries persist in the metallic states. Beyond this, the waters deepen. More experimental landmarks than are currently available may be needed to make confident theoretical progress.

Part of the theoretical challenge posed by interactions in MATBG flows from the topology[43,63,76,124-126,130,131] of the flat bands, which is non-trivial and excludes the possibility of constructing simple accurate generalized-Hubbard tight-binding model representations of MATBG physics. Lattice models that have been derived[75,79,131,132], like the one illustrated in Figure 6a, make a difficult compromise between accuracy and orbital proliferation. A related complication is that the filled remote valence bands play a role even when they are spectrally isolated and inert. To understand why, we first recognize that the position and sublattice dependence of the flat band wavefunctions has a strong and complex dependence on momentum in the moiré Brillouin-zone – something that would not be the case if the system were described by a simple tight-binding model. Electrostatic[133] and exchange[105] interactions with the remote bands, which have a deficiency in charge near AA positions in the moiré pattern, radically change the shape of the flat bands as they are filled because of relative shifts in the mean-field energies of states at different momenta. For example, the valence band minimum, which is invariably located at the Γ-point in the moire Brillouin-zone when interactions are neglected as illustrated in Figure 6b, has less weight at AA positions and is therefore pushed up relative to Γ, possibly by enough to make it the valence band maximum. Broken rotational symmetry, discussed above, moves the Dirac points away from the Brillouin-zone corners, and additional broken symmetries can gap the Dirac points. Some plausible theoretical models[134] for the Fermi surface shape from which the strongest superconductivity emerges and band fillings near -2.3 are illustrated in Figure 6(c-f). It may be that the band energy scale is small enough relative to the interaction scale that band dispersion is nearly irrelevant. However, if the details of the band dispersion do indeed play a critical role it appears that help from experiments that probe Fermi surface shape is sorely needed.

Progress is being made[135] toward accurate measurements of Fermi surface sizes and shapes thanks to a combination of advances in ARPES technology and sample preparation. These measurements will be of critical importance to test the reliability of theoretical models. Beyond this, one key challenge for theory is the identification of the mechanism responsible for the rapid increase in resistivity at low temperatures found in experiment which points at a minimum to a surprisingly high density of low-energy inelastic scatterers. As mentioned above the possibility that these observations might be explained by electron-phonon scattering[136] has not yet been excluded. Finally, we come to the serial enigma of strongly-correlated electron physics, superconductivity. What is the source of the attractive interaction that pairs electrons in MATBG flat bands over wide ranges of band-filling? Phonon-mediated attraction must, and has been,[62,83,137,138] considered as a candidate, along with other mechanisms[82,139-141]. The recent observation[142] that moving the electrical gates used to vary band filling factor, which act as screening planes, closer to the bilayer favors superconductors over other states in the MATBG phase diagram provides an important constraint on theoretical possibilities and may suggest that electron-electron interactions do not play the key role.

**Outlook**

The most important obstacle to immediate progress in understanding the properties of twisted multilayers formed from graphene and other two-dimensional semiconductors or semimetals is achieving more reliable control over average twist-angle, and twist-angle disorder. We now know that uncontrolled strains related to details of how mechanical contact is established between the twisted layers can strongly influence physical properties, including for example basic features of the phase diagram which expresses the appearance or absence of superconductivity, orbital magnetism, and gapped states. The more perfect the twist angle control, the better the periodicity of the moiré pattern, and the richer the electronic phase diagram. Progress is being made in this direction, and we can expect it to continue. With better twist angle control the influence of other control parameters on physical properties can be more reliably assessed. For example it will be desirable to study how properties vary with i) transverse electric fields due to unbalanced gating which is expected to induce bulk gaps populated by chiral states localized along domain walls between AB and BA regions, ii) hexagonal boron nitride substrate orientations that explicitly break the sublattice symmetry of adjacent layers, and iii) the separation between samples and metallic gates, which will influence the strength of Coulomb interactions between electrons, and iv) carbon isotope effects, which will influence the strength of phonon-mediated interactions.

As in untwisted graphene, bilayers are not the whole story. For example trilayers[29] can have larger magic twist angles which mitigate some of the most severe experimental challenges and increase energy scales, including those of superconducting and magnetic transition temperatures (Figure 7a). An alternative approach to create robust flat bands which do not require fine tuning by imposing an external periodic potential[143], could also be explored. This strategy could in principle extend the range of correlated electron physics to larger systems and higher temperatures. The emergence of such flat bands was demonstrated[9] in a graphene membrane that underwent a buckling transition, resulting in a strain-induced periodic pseudomagnetic field (Figure 7b). An entirely new playground opens in tiny-angle TBG, where the significantly larger moiré cells host a topologically non-trivial network of one-dimensional channels formed at sharp boundaries between AB and BA domains[144-148] (Figure 7c,d). Other two-dimensional materials families are also promising. At the current stage of progress in device fabrication, the transition metal dichalcogenide 2D electron systems seem particularly promising. These have many-body strong correlation physics that is on its face more similar to that found in some atomic crystals. Recent studies reported experimental evidence for Mott insulator physics in the triangular moiré superlattices formed from transition metal dichalcogenides[30,33], and demonstrated some of the promise of moiré quantum simulation by exploring properties as a function of band filling in a way that is not possible in atomic crystals[30]. Because of the possibility of tuning carrier density over broad ranges without introducing disorder, we can be sure that studies of moiré superlattices will advance many-particle physics. Although we cannot be certain at the moment what the implications will be for our understanding of other condensed matter systems, the history of science teaches us to be optimistic.

**Figure captions**

**Figure 1. Structure of monolayer graphene and twisted bilayer graphene**. a) Honeycomb lattice of graphene with the two sublattices, A and B, represented by white and blue circles. **a$_1$**, **a$_2$** are the honeycomb lattice vectors and **a** is the Carbon-Carbon distance. Adapted from reference[35]. b) Graphene's hexagonal Brillouin zone (shaded area) showing the two inequivalent sets of *K* (full circles) and *K'* Brillouin zone corners (open circles). **G$_1$** and **G$_2$** are the reciprocal lattice vectors. Adapted from reference[35]. c) Low energy band structure of graphene showing the two inequivalent Dirac cones and the corresponding Dirac points located at the *K* and *K'* corners of the Brillouin zone (red hexagon). The pink plane defines the Fermi surface at charge neutrality. Adapted from reference[35]. d) Moiré pattern formed by two superposed honeycomb lattices. The bright spots correspond to local AA stacking where each atom in the top layer has a bottom layer atom roughly directly underneath it. In the darker, AB (Bernal) stacked regions, each top-layer atom in the A sublattice sits roughly above a B atom in the bottom layer, while top-layer B atoms have no partner in the bottom layer. e) Superposing the Brillouin zones of the top (green) and bottom graphene layers gives rise to the two hexagonal mini-Brillouin zones (black hexagons) of the moiré superlattice. The latter are spanned by the *K$_1$*-*K$_2$* and *K$_1$'*-*K$_2$'* segments connecting adjacent corners of the original Brillouin zones of the bottom and top layers respectively. Adapted from reference[20]. f) Dirac cones of top (green) and bottom (red) graphene layers showing their twist angle dependent displacement, $k_\theta$, and intersection points. Adapted from reference[16].

**Figure 2. Electronic properties of twisted bilayer graphene**. a) Top: schematics of a scanning tunneling microscopy (STM) measurement of a twisted bilayer graphene (TBG) sample deposited on an hBN substrate. The stack is supported by a Si/SiO$_2$ backgate used to control the Fermi level by electrostatic doping. Bottom: Topography of TBG with a twist angle of $1.8^0$, grown chemical vapor deposition (CVD) shows a uniform moiré pattern with period 7.5 nm and no observable twist-angle inhomogeneity. Adapted from reference[14]. b) Bottom trace: dI/dV spectrum at the center of an AA region of the moiré pattern in panel *a* shows two van Hove singularity (VHS) peaks flanking the Dirac point and separated by an energy of 85mV. Top trace: dI/dV spectrum at the center of an AA region in a magic-angle TBG ($\theta \sim 1.16^0$) where the merged VHSs produce a DOS peak with a small (12meV) pseudogap at the Fermi level (zero sample bias), suggesting the emergence of a correlated state. Adapted from reference[14]. c) Renormalized Fermi velocity as a function twist angle obtained from Landau level spectroscopy (symbols) compared to a theoretical simulation (solid line). Adapted from reference[16]. d) Renormalized Fermi velocity as a function twist angle in the range $0.18^0 < \theta < 1.2^0$, shows a sequence of magic twist-angles, where the Fermi velocity vanishes. The twist angle is expressed in terms of the dimensionless parameter $\alpha = w/\hbar v k_\theta$, where *w* is the interlayer tunneling strength and $k_\theta \sim 2K \sin(\theta/2)$ as defined in the text. Adapted from reference[47]. e) Band structure of MATBG shows the formation of a narrow band near charge neutrality upon the merging of the two saddle points when approaching the magic angle. The flat-band gives rise to the narrow peak shown in the top trace of panel (b). Adapted from reference[21].

**Figure 3. Comparing the phase diagram of MATBG and high temperature superconductors.** a) Schematics of experimental setup for transport measurements on TBG samples encapsulated in hBN. Adapted from reference[20]. b) Phase diagram showing the dependence of the superconducting critical temperature on carrier density in the partially filled flat-band. Two dome shaped superconducting regions (blue) are observed flanking an insulator (red) at filling ν = -2. Adapted from reference[21]. c) Schematic cuprate phase diagram, showing two superconducting domes (green) flanking an antiferromagnetic insulator phase (orange), a pseudogap and nematic phase (purple), a strange metal phase (white) and a normal metal (blue). Adapted from, Holger Motzkau, Wikimedia Commons, 2013. d) Carrier dependence of conductance in magic angle TBG (MATBG) shows pronounced minima at several integer fillings. Adapted from reference[20]. e) Density dependence of $T_c/T_F$ ($T_c$ superconducting temperature, $T_F$ Fermi energy) as a function of doping for magic-angle MATBG (red filled circles) compared to other strongly correlated systems marked by the horizontal dashed lines. Adapted from reference[21].

**Figure 4. Flat band magic angle TBG observed with local probes.** a) Twist-angle disorder and strain are quantified using a combination of Raman spectroscopic and low-energy electron diffraction imaging. The twist angle between TBG layers varies on the order of $2^0$ within large (50-100 μm) single-crystalline grains, resulting in changes of the emergent Raman response by over an order of magnitude. Rotational disorder comes about by variations in the local twist angles between differing contiguous subgrains,~1 μm in size, that themselves exhibit virtually
no twist angle variation The color scale represents the twist angle variation. Adapted from reference[86]. b) STM topography of tear-and-stack TBG reveals large twist-angle inhomogeneity. The bright spots correspond to AA regions of the moiré-pattern. The green hexagon represents a region where the twist angle is ~ $1^0$. (E.Y. Andrei and Y. Jiang unpublished). c) Twist-angle disorder detected with a scanning nano SQUID-on-tip. Adapted from reference[89]. d. dI/dV spectra in the center of an AA region of an MATBG prepared by the tear-and-stack method with twist angle θ ~ 1.07° at several fillings of the flat band. Fully occupied band ($V_g$ = +55 V) corresponds to the red trace; empty band ($V_g$ = -55 V) to the blue trace and partially filled band ($V_g$ = 0 V) to the green trace. The spectrum of the partially filled band features a pseudogap that splits the DOS peak. The two remote bands flanking the flat band are marked by the blue arrows. Inset: dI/dV spectrum on an AB site taken at $V_g$ = +55 V. The Fermi level for all spectra is at zero bias voltage. Adapted from reference[24]. e) Gate voltage (filling) dependence of the dI/dV intensity measured by STS at the Fermi level in the center of an AA site in a MATBG sample with twist-angle θ ~ $1.06^0$. The spectra show clear dips at integer fillings marked by shaded bars, $\nu = 0, \pm 1, \pm 2, \pm 3, \pm 4$, mirroring the results obtained in similar samples by using transport measurements. Adapted from reference[24]. f) Twist-angle dependence of low temperature inverse compressibility in TBG measured in tear-and-stack samples with twist angles as marked at 12T, shows peaks at integer fillings that become more pronounce upon approaching the magic angle at θ = $1.07^0$. Adapted from reference[69].

**Figure 5. Nematicity, orbital magnetism and strange metal behavior in magic angle TBG.** a) Global nematic charge order observed in the partially filled flat-band using STS based charge spectroscopy. Alternating negative (red) and positive (blue) charge stripes are aligned roughly

along a crystal axis of the moiré-pattern. Figure adapted from reference[24]. b) Superconducting phase diagram in MATBG shows several correlated insulators flanked by four superconducting domes. Adapted from reference[22]. c) Phase diagram of MATBG includes competing phases with superconducting and nematic order as marked. Adapted from [149]. d) Hall density, $n_H$, as a function of moiré cell filling, $v$, measured in a low magnetic field, $B = 0.8T$. Lifshitz transitions associated with the formation of van Hove singularities and spectral gaps at integer fillings $\pm 2, \pm 3$ are seen as diverging $n_H$. Here $n$ is the gate induced carrier density and $n_0$ the carrier density to fill one moiré cell. Insets: schematic evolution of the Fermi surfaces (red triangles) from separate pockets centered on the corners of the mini-Brillouin zones for low doping (bottom) and their merging together near the VHS (top) close to $v = 2$. Adapted from reference[66]. e) Evolution with magnetic field of the symmetrized Hall resistance, $\bar{R}_{xy}$, as a function of reduced filling , $(n/n_0 - v)/B$, shows the emergence of Chern insulators revealed by the well-quantized Hall plateaus at fillings $v = 3$ (top) and $v = 2$ (bottom) with Chern numbers $C = -1$ and $C = -2$ respectively. Adapted from reference[66].

**Figure 6. Moiré bands and Fermi surface.** a) Moiré bands implied by a ten-band tight-binding model for the valley projected bands of magic-angle twisted bilayer graphene. b) Energy *vs*. momentum for the flat conduction and valence bands. Notice that the flat valence band minimum is at Γ and that there are two linear band-touching Dirac points at the Brillouin-zone corners K and K'. (After Figure2 of reference[132] ). These band models are constructed from electronic structure calculations performed at charge neutrality. c-f) Model Fermi surfaces of magic-angle twisted bilayer graphene at band filling $v = -2.33$ near where the strongest superconductivity is observed. Because the magic-angle bands are so flat they are very sensitive to details, including band renormalization by electrostatic and exchange interactions. The band energies in this panels are indicated by the color scales at right, and are based on the continuum model calculations in Reference[134] which account for strain and for interactions at the mean field level. Occupied states are indicated by enclosed momentum space pixels. The Fermi surfaces in (c) and (d) are based on restricted mean-field calculations in which spin/valley flavor symmetries are unbroken and differ in the interaction strength parameter $\varepsilon^{-1}$ used to account for environmental (gate plus dielectric) screening of the Coulomb interaction. The Fermi surfaces in (c) and (d) both exhibit nematic order, which occurs nearly universally in mean-field band calculations. The band state at Γ is unoccupied at this band filling in (d) ($\varepsilon^{-1} = 0.08$), even though it is at the band minimum when interactions are neglected, but occupied in (c) ($\varepsilon^{-1} = 0.04$). In (e) (($\varepsilon^{-1} = 0.04$) and (f) ($\varepsilon^{-1} = 0.08$), flavor symmetries are broken so that two of the four bands are empty and two are nearly filled, and irregularly shaped hole-like Fermi surfaces are seen near the Γ point in the Brillouin-zone. Figure credits: AH MacDonald and M. Xie, unpublished.

**Figure 7. Reaching beyond MATBG.** a) Transport measurements on trilayer ABC graphene reveal a superconducting state emerging at low temperature close to ¼ filling. Adapted from reference[29]. b) STM topography of a buckled graphene membrane supported on $NbSe_2$ reveals a triangular lattice of alternating crests and troughs resulting in a periodic pseudo-magnetic field with maximum amplitude ~120T. Adapted from reference[9]. c) Network of 1D channels at the boundary of AB and BA domains in tiny angle TBG imaged with Piezoresponse Force Microscopy (PMF). Adapted from reference[144]. d) Same as (c) imaged with TEM. Adapted from reference[148].

**Acknowledgements:** EYA acknowledges support from DOE (DOE-FG02-99ER45742) and the Gordon and Betty Moore Foundation (GBMF9453), AHM acknowledges support from DOE BES grant DE- FG02-02ER45958 and from Welch Foundation grant TBF1473.

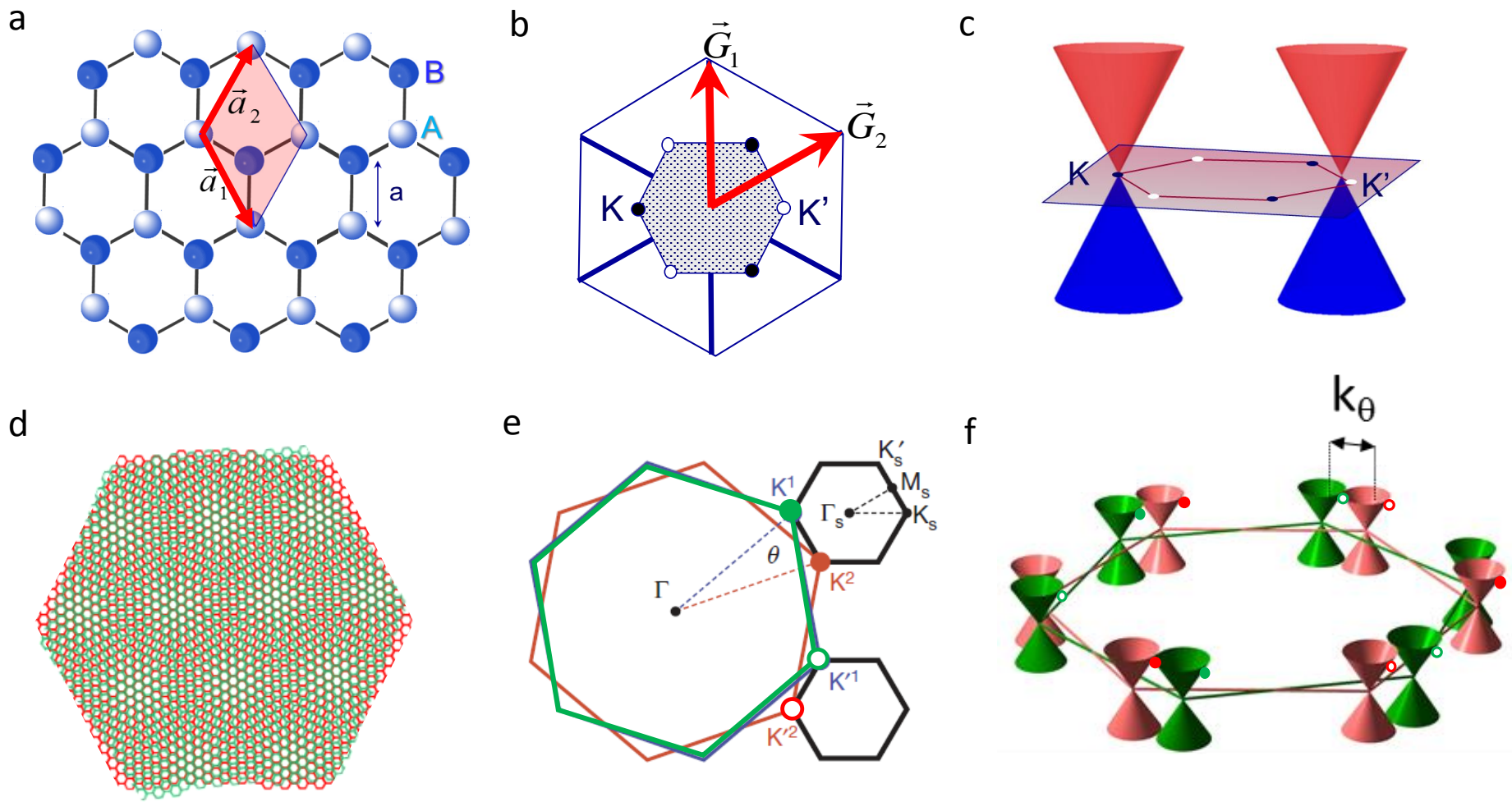

**Figure 1**

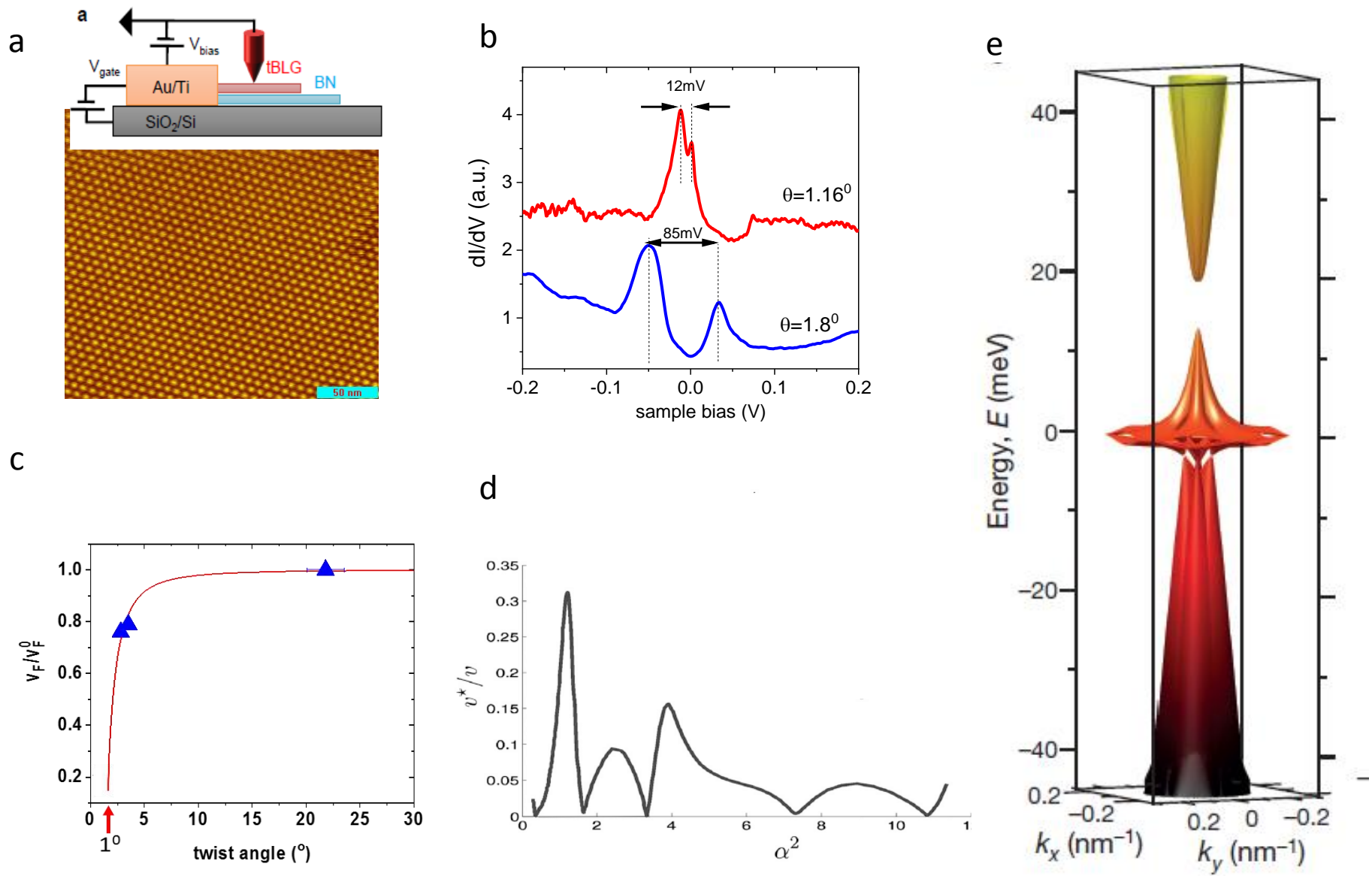

**Figure 2**

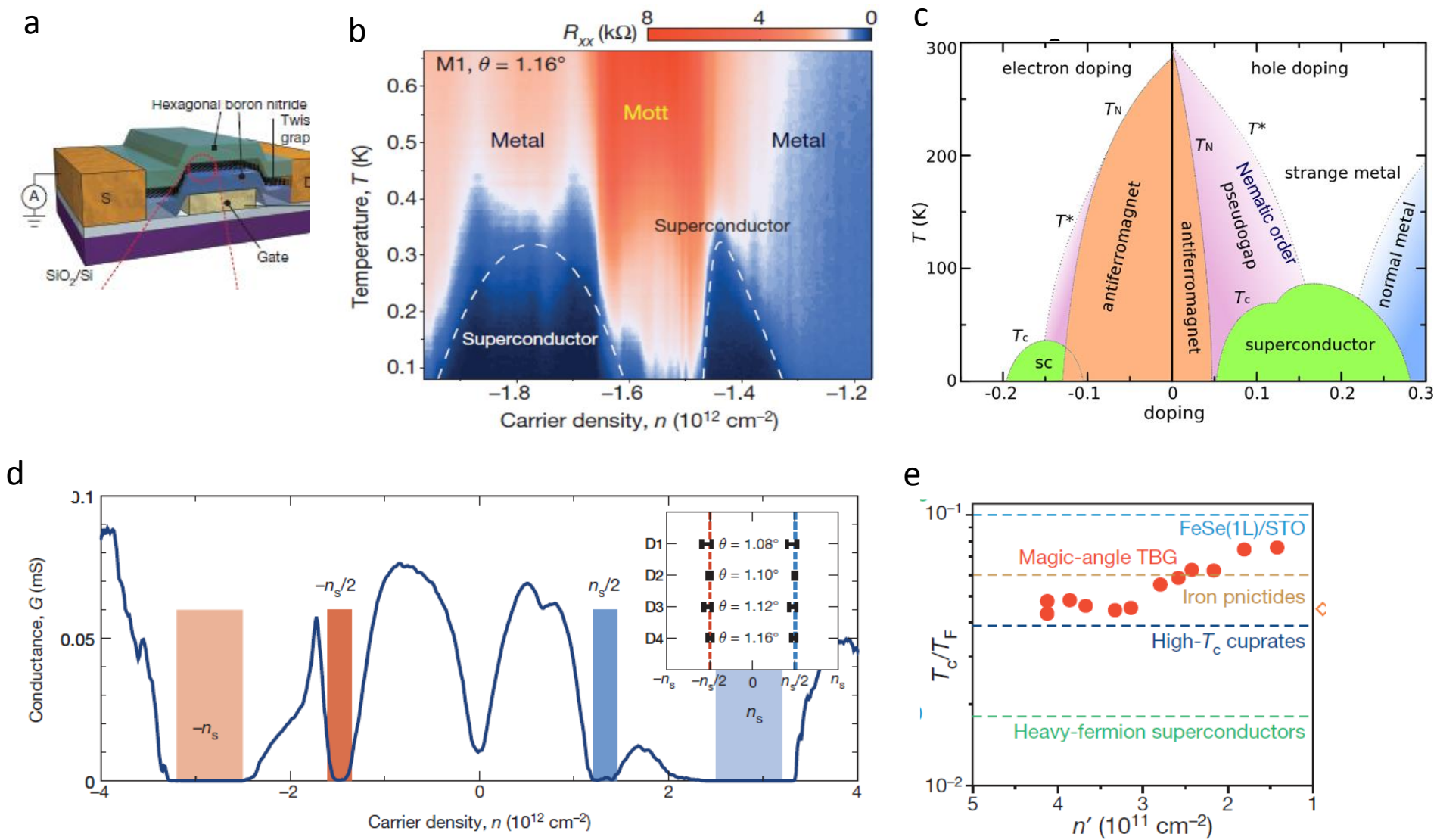

**Figure 3**

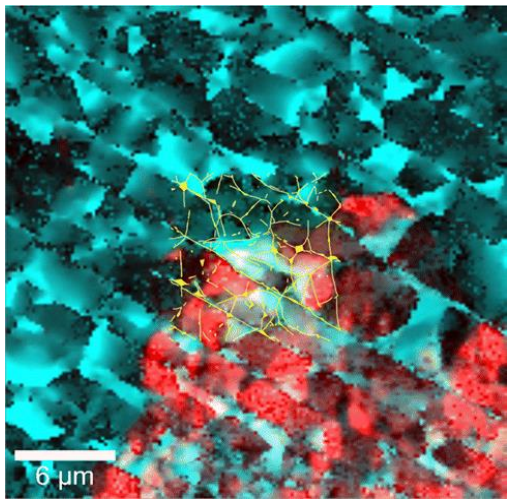 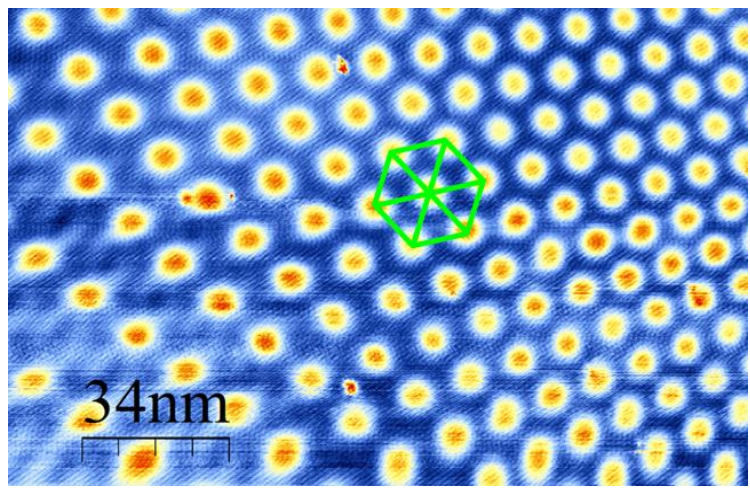 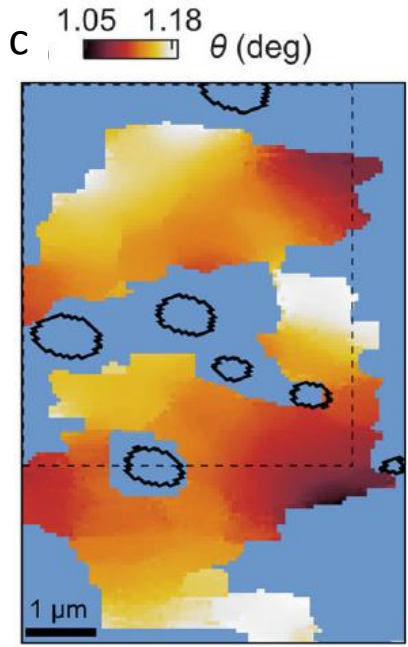

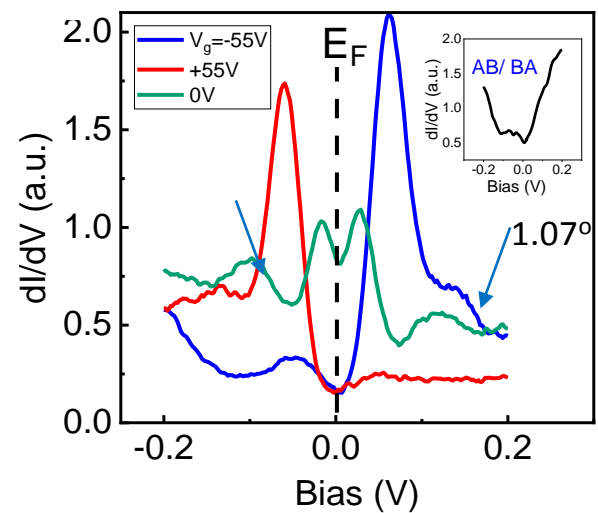 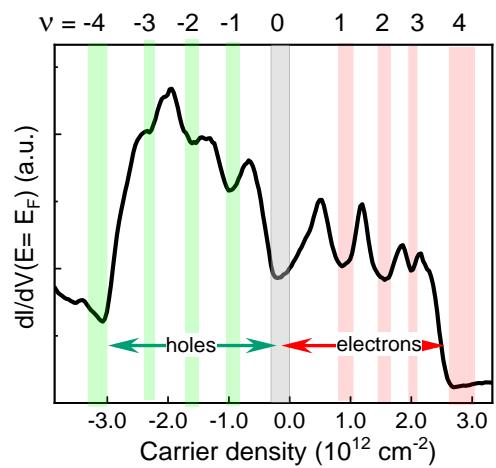 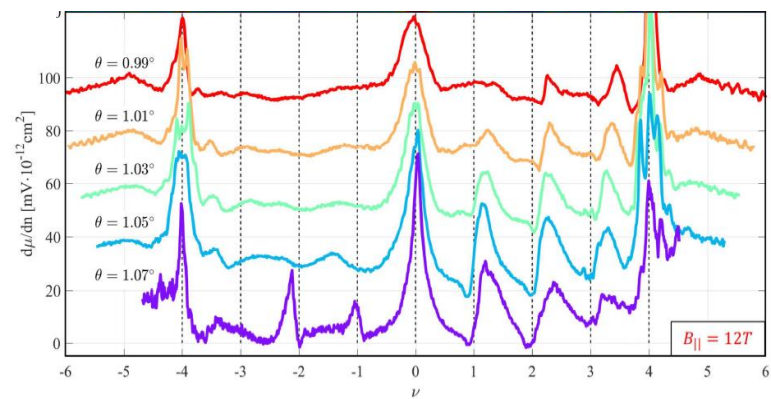

**Figure 4**

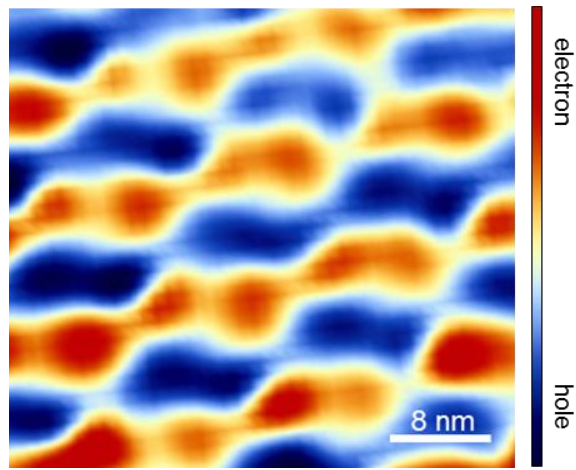
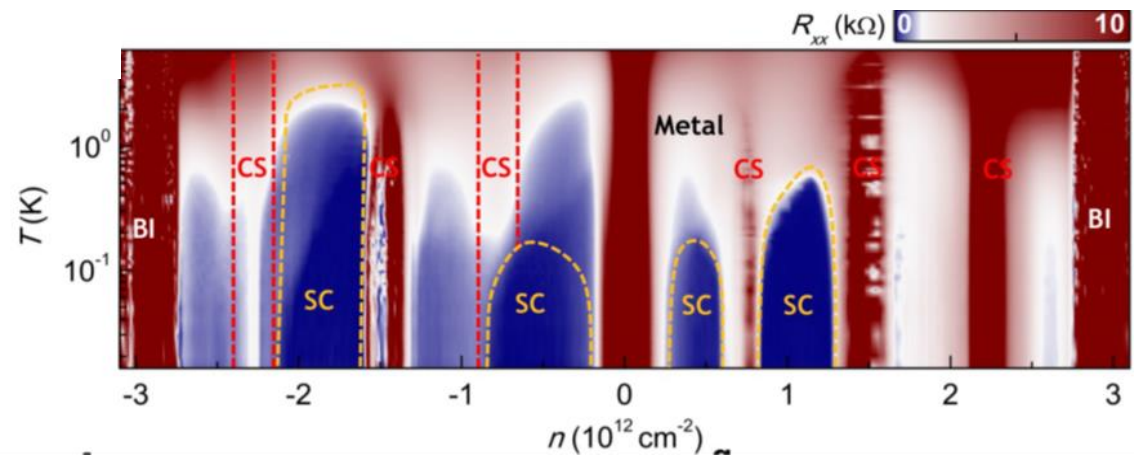
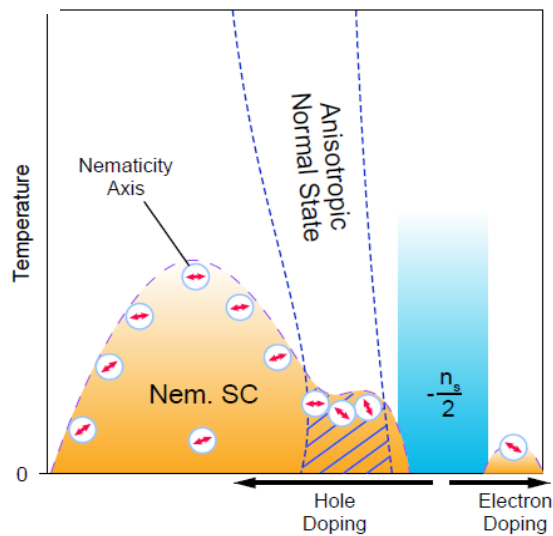
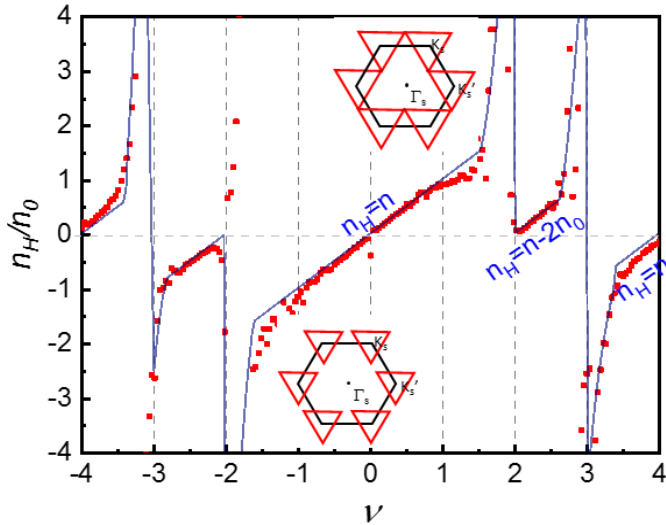
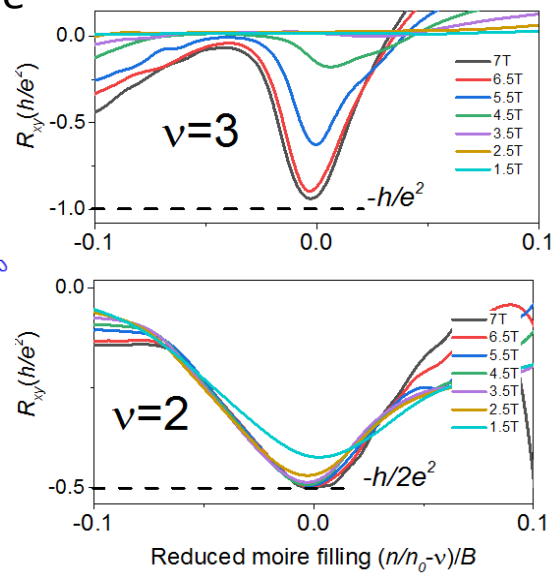

**Figure 5**

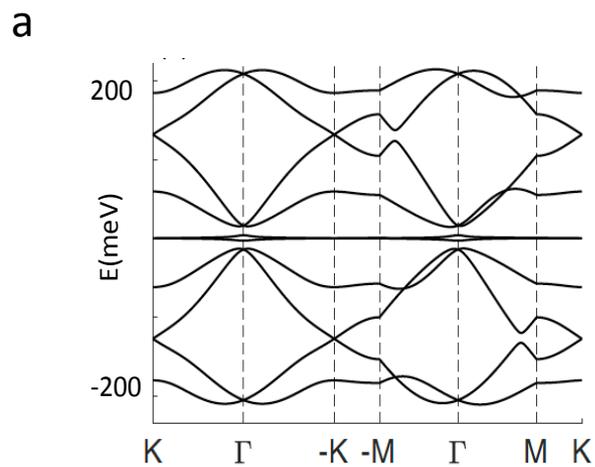
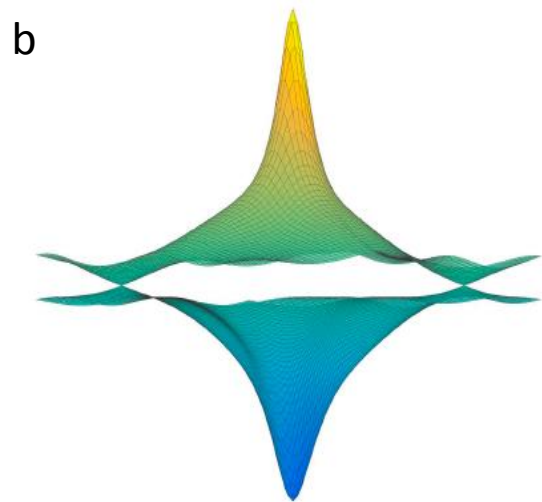
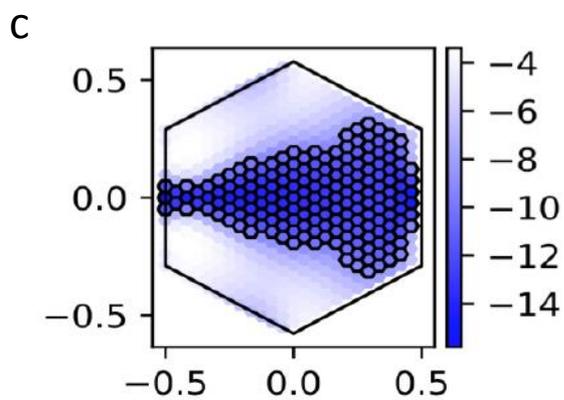
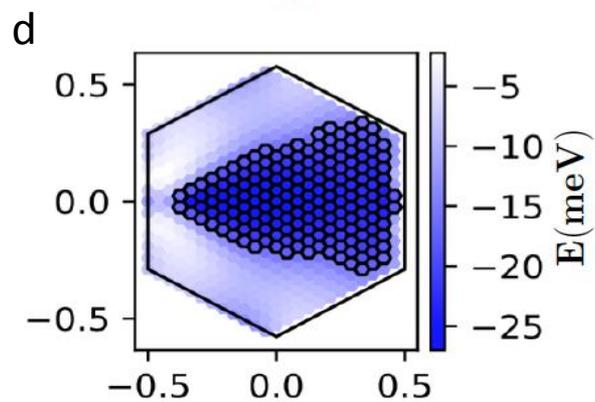
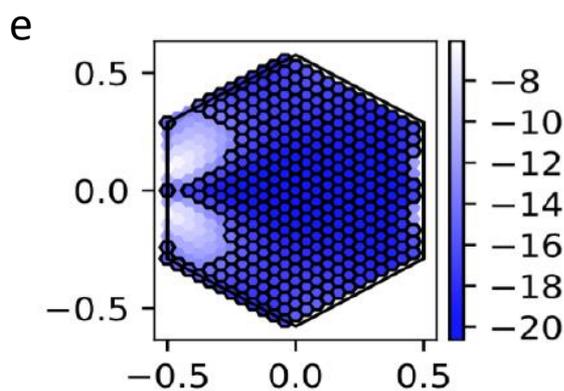
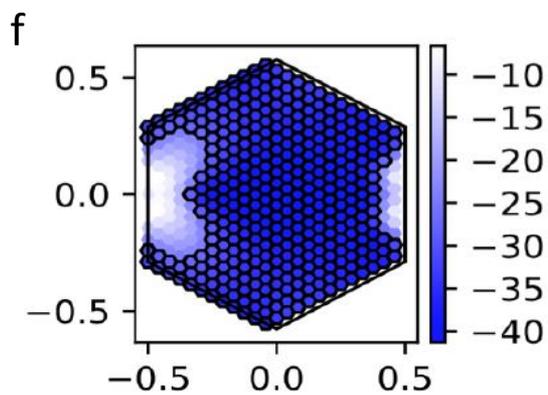

**Figure 6**

a 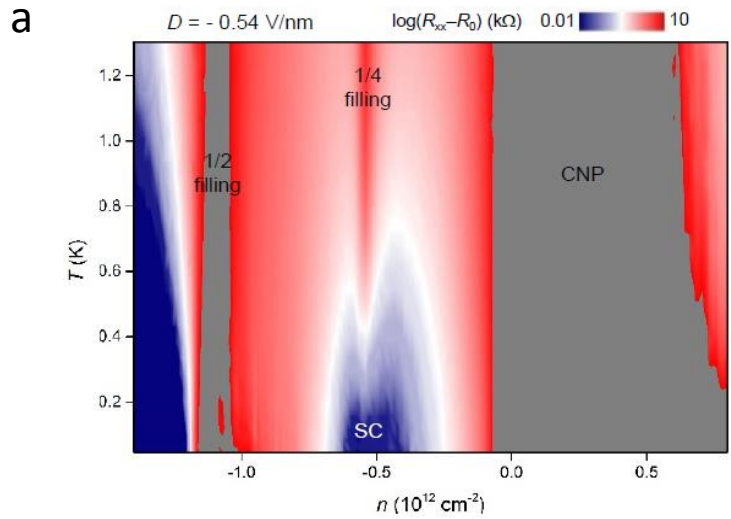 b 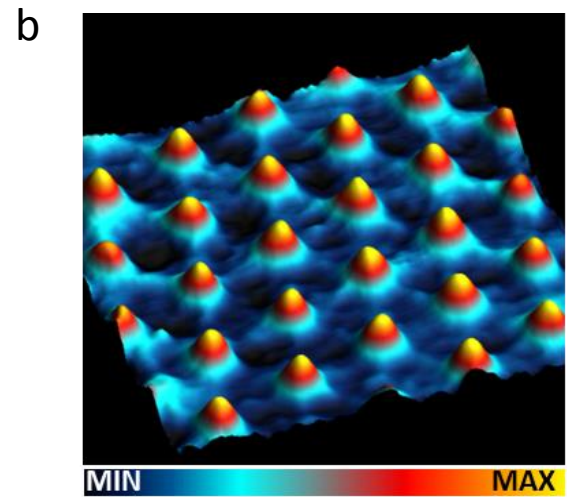

c 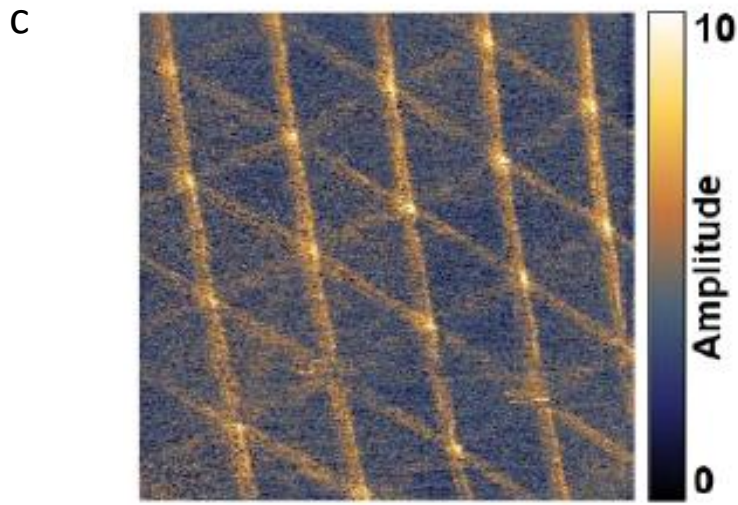 d 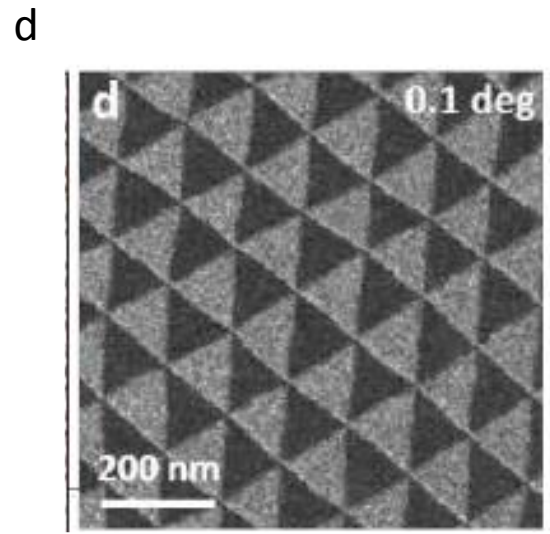

**Figure 7**